# Grain Boundary Segregation and Embrittlement of Aluminum Binary Alloys from First Principles

Nutth Tuchinda[a*], Gregory B. Olson[a], Christopher A. Schuh[a, b]

[a]Department of Materials Science and Engineering, Massachusetts Institute of Technology, 77 Massachusetts Avenue, Cambridge, MA, 02139, USA

[b]Department of Materials Science and Engineering, Northwestern University, Clark Street 633, Evanston, IL, 60208, USA

*Correspondence to nutthtu@mit.edu

## Graphical Abstract

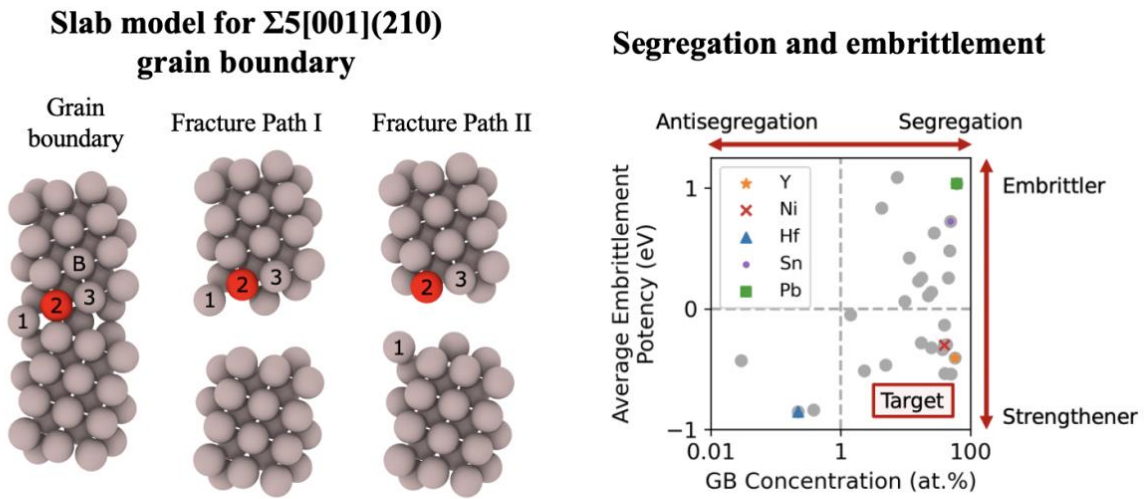

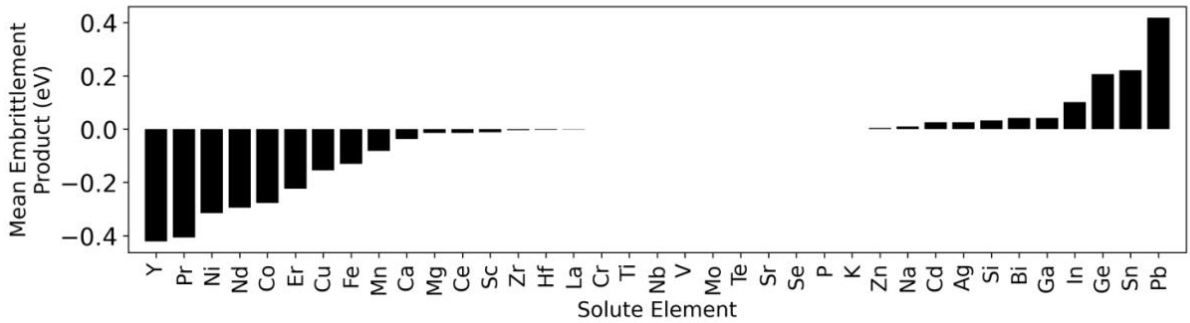

**Abstract**

Grain boundary segregation controls properties of polycrystalline materials such as their susceptibility to intergranular cracking. It is of interest to engineer alloy chemistry to enhance grain boundary cohesion to prevent intergranular failure. While there is collectively a large first-principles dataset for grain boundary embrittlement in multiple Al-based binary alloys, the methodologies used for the first principles calculations, as well as the analyzed fracture paths, are variable amongst studies. Here, we reevaluate and compute grain boundary segregation and embrittlement from all-electron first-principles for the Σ5[001](210) Al grain boundary. We explicitly evaluate multiple fracture paths, and provide a study case of the chemical trends of the preferred fracture paths across 69 binary Al alloys. The results suggest that neglecting certain low energy fracture paths can lead to errors of estimating embrittlement potency up to the order of 1 eV per solute atom, especially for multiple d-block transition metal solutes that are of engineering interest. The database calculated here also permits a comprehensive comparison between all-electron and pseudopotential methodologies. The effects of Hubbard $U$ density functional theory on grain boundary segregation and embrittlement in Al(Sc) are found not to be significant in terms of the relative energetic calculations of grain boundaries and free surfaces (differences are of order 0.1 eV or less).

## 1. Introduction

Grain boundaries (GB) are ubiquitous in engineering alloys, and can control critical properties such as fracture [1–7] and the stability of nanostructured alloys [8–13]. GBs comprise numerous atomic sites that provide solute elements with lower energy states vis-à-vis bulk sites due to the introduced elastic strain and chemical bond changes [14–17]. The resulting grain boundary segregation affects the boundary energy, and at the same time can dramatically influence the tendency for intergranular fractures [18–20]. Therefore solute species with a tendency to segregate but not embrittle grain boundaries should generally be selected in an alloy design process [21] to avoid intergranular fracture in engineering alloys [3,22–24].

There have been numerous studies of grain boundary segregation and embrittlement in Al-based alloys [18,19,25–29]. For instance, the effects of B, Zn, Na and Mg on GB cohesion are demonstrated by Zhang *et al*. in Refs. [25–28,30] with full linearly-augmented plane wave density functional theory (FLAPW DFT) on Σ5[001](210) GBs. A more recent comprehensive analysis of multiple boundaries is provided by Mahjoub *et al*. through projector augmented wave method (PAW) DFT calculations [18] throughout numerous transition metal solute elements on multiple coincident site lattice boundaries, including the same Σ5[001](210). However, past literature on GB embrittlement as reviewed and summarized by Lejcek *et al*. [31] and Gibson and Schuh [19] suggest a large variability of the data provided by ab initio modeling which may arise due to the inconsistency of the methods and parameters used [32]. For instance, the reported range Al(Sc) and Al(Zr) segregation energies at Σ5[001](210) spans more than 1 eV [18,33,34]. There could also potentially be multiple fracture paths, each with different embrittlement energetics, which are not systematically covered in these calculations [6,27].

Recent computational advances allow FLAPW calculations in a high-throughput manner, thereby without the reliance on construction of a pseudopotential. This creates an opportunity for a robust benchmarking of GB segregation and embrittlement across the periodic table for binary Al alloys. In this work, we employ FLAPW and PAW DFT methods to study grain boundary segregation and embrittlement potency of Al-based binary alloys using Σ5[001](210) grain boundaries for 69 substitutional alloys (from Na to Bi). Multiple fracture paths are also considered for all these binary alloys, with significant variability uncovered as a result. In addition to providing a large, self-consistent first-principles database for GB segregation and embrittlement in Al, these results speak to the need of a statistical perspective [6] of GB fracture in general polycrystals.

## 2. Grain Boundary Calculations from Density Functional Theory

The FLAPW calculations [35–43] in this work are performed following the methodology described in Refs. [25–28] with the fleur DFT MaX-R6.2 software package [44] using thin film boundary conditions [39,45]. In essence, a Perdew-Burke-Ernzerhof generalized gradient approximation functional [46] is used with the planewave cut off $K_{max}$ = 4.5, $G_{max}$ = 15 and $G_{max,ec}$= 12.5. The systems are treated non-magnetically. Histogram Brillouin zone integration is used with σ = 0.0025 Ha over the k-point mesh of 7×7. The muffin-tin radius of Al is chosen as 2.3 a.u., with the rest for the solute elements listed in the supplemental material. The spherical harmonic and nonspherical expansion are also kept up to $l$ = 8 and $l_{nonsph}$ = 6 for Al, and $l$ = 12 and $l_{nonsph}$ = 8 for the solute species. The density convergence is assumed when the charge density differences are less than $10^{-6}$ e/(a.u.)$^3$. Force relaxation is conducted for all configurations until the force norms are all less than 0.01 eV/a.u. PAW calculations done in this work are conducted with the Fermi smearing method and identical k-point grid. The pseudopotentials used here are tabulated in the supplemental material. Analyses in this work are done through OVITO and related Python software packages [47–58].

In this work we construct a 25-layer slab Σ5[001](210) GB [18,25–28,59,60] similar to Zhang et al. [25–28] using the gb_code [61] package, which we present in Fig. 1 for the corresponding segregation sites (defined as site 1, 2 and 3) and the reference bulk site (site B). A vacuum region is added to both sides of the GB and free surface slab with a thickness of at least 4.5 a.u. or with $d$ and $\tilde{d}$ of 1.5 and 4.5 a.u. respectively beyond the slabs. The calculated GB, free surface and fracture energy from the slab model are listed in Table 1, all of which are in very good agreement with previously reported values [26]. We note that we define two fracture paths as path 1 and 2 for segregation site 2 and 3 in Fig. 1. As we shall see later, certain solute species may prefer to fracture with or without Al near to the solute site [27].

**Table 1** Calculated grain boundary energy, free surface energy and fracture energy of the modeled Σ5[001](210) grain boundary in J/m$^2$.

| Structure | Ref. [26] | This work |
| --- | --- | --- |
| Grain boundary energy | 0.501 | 0.493 |
| Free surface energy | 1.016 | 1.015 |
| Fracture energy | 1.531 | 1.538 |

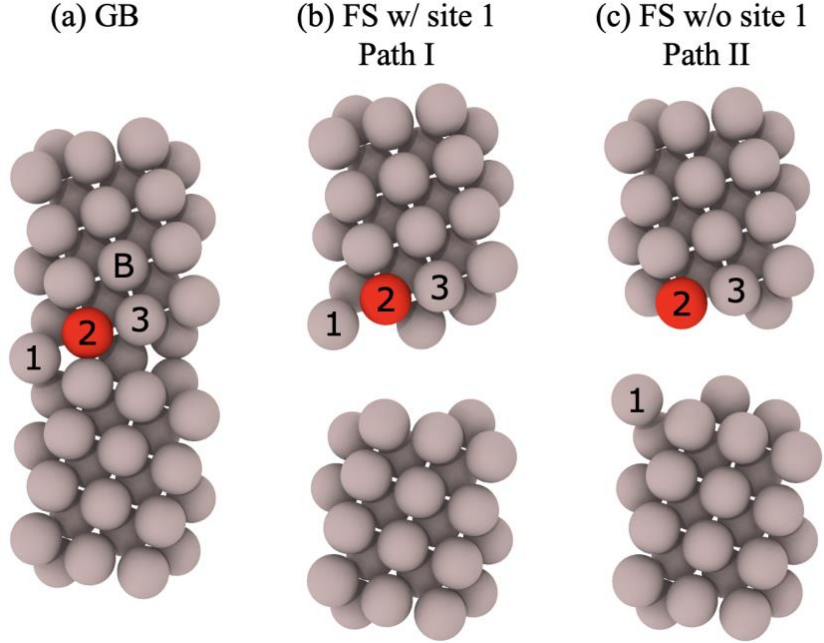

**Fig. 1** (a) Σ5[001](210) grain boundary structure used in this work with an example of a solute at site 2. Here we consider two fracture paths for segregated solute species at site 2 and 3 as shown in (b) and (c) for the solute free surface slab (FS) with and without site 1 respectively.

### 3. Grain Boundary Solute Segregation

Here, we define segregation energy ($\Delta E_i^{\text{seg}}$) as the energetic difference between a system with a solute occupying a GB site ($E_i^{\text{GB}}$) referenced to that with a single solute atom in a bulk site ($E^{\text{Bulk}}$):

$$\Delta E_i^{\text{seg}} = E_i^{\text{GB}} - E^{\text{Bulk}} \tag{1}$$

With the local site occupation probability or concentration derived from a classical isotherm as [62,63]:

$$\frac{X_i^{\text{GB}}}{1 - X_i^{\text{GB}}} = \frac{X^{\text{C}}}{1 - X^{\text{C}}} \exp\left(\frac{-\Delta E_i^{\text{seg}}}{RT}\right) \tag{2}$$

where $X_i^{\text{GB}}$ and $X^{\text{C}}$ are the local concentration at site type "$i$" and in the bulk respectively, and $k_B$ and $T$ are Boltzmann constant and temperature. Negative segregation energy sites give positive values on the right-hand side of Eq. (2), and are therefore preferential sites for solute segregation.

We show an overview of grain boundary segregation energy of all 3 sites for every solute element considered in the periodic table plot of Fig. 2. As an example, in Al(Zn) we compute the most segregating site to be site 2 with a segregation energy of −0.19 eV (vis-à-vis, also, −0.19 eV in Ref. [27]). Sites 1 and 3 also show a good match to the prior literature at −0.01 and −0.08 eV vis-à-vis −0.01 and −0.09 eV respectively.

Next, we look at systems that are known to segregate experimentally. We take Al(Au) as an example; site 1, 2 and 3 show segregation energies of −0.064, −0.507 and −0.207 eV respectively. These numbers translate at $X^{\text{C}} = 0.01$ at.% and $T = 573$ K to an average grain boundary concentration of ~12 at.% (which is an enrichment ($X^{GB}/X^C$) of ~$10^3$) vis-à-vis experimentally-reported value at the order of up to ~5 at.% for high-angle GBs [64]. Zn, Mg and Cu in 7000-series Al alloys are also known to be segregating to random high-angle GBs [65,66]. As an example from Ref. [65], we apply $X^C_{Zn} \approx 2.2$ at. %, $X^C_{Mg} \approx 2.8$ at. % and $X^C_{Cu} \approx 1.4$ at. % at $T \approx 800$ K (the solutionizing temperature) to obtain $X^{GB}$ of 16, 30 and 37 at.% (compared to the reported values of 4, 5.3 and 2.3 at.% respectively). While the exact values should probably not be compared in this case (as the ternary alloy obviously would have site competition and interactions amongst the solutes which we do not consider in our model [67–70]), it is encouraging that all three solutes are expected to segregate as is indeed experimentally observed. It is also important to remember that such concentrations are not well defined thermodynamic quantities by themselves, as they result on the choice of a GB thickness definition [71–73]. What is more, the site fractions from Σ5[001](210) very poorly compare with the GB sites in polycrystals [63,74] upon which experiments are performed.

We further summarize the periodic trend in Fig. 3 using a solute concentration of 1 at.% and $T = 700$ K to calculate average grain boundary concentrations, and also note the most segregating sites using the symbol (circle for site 1, triangle for site 2 and star for site 3). Overall, there are numerous d-block metals of interest, such as V, Nb and Ta that should, in principle, improve grain boundary cohesion due to their strong cohesive energy relative to Al [3], although in this case the potency is limited by their tendency of solute segregation. To directly obtain embrittlement potency on top of the segregation energies in Fig. 2, we proceed next to evaluate embrittlement potency using the slab model.

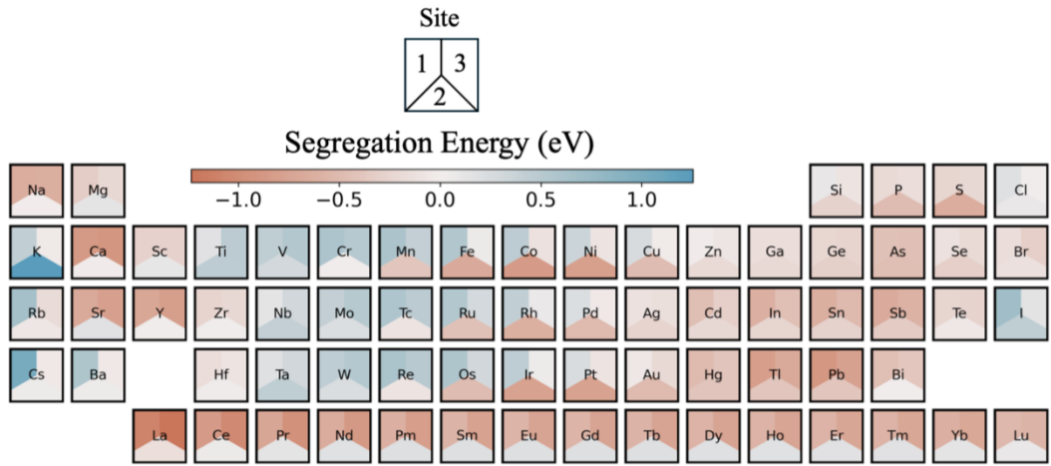

**Fig. 2** A periodic table plot of segregation energy for all calculated substitutional alloys (see supplemental Table S1 for numerical details).

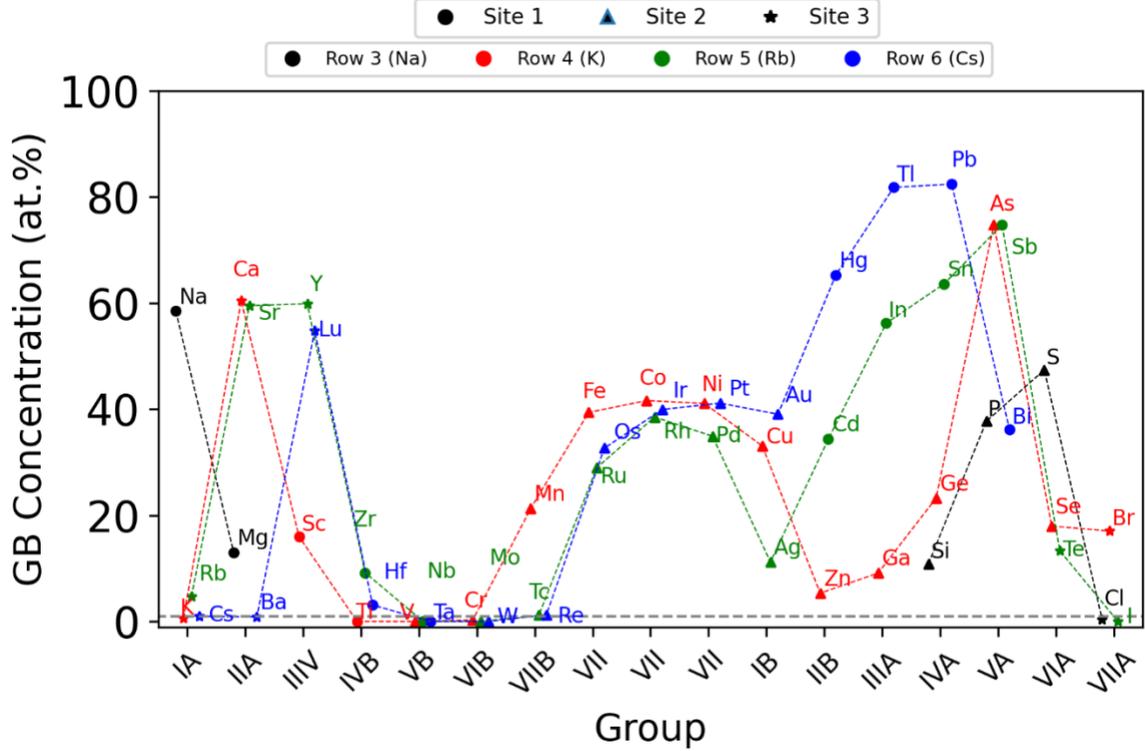

**Fig. 3** Periodic table plots for grain boundary concentration calculated at $T$ = 700 K and $X^{\text{tot}}$ = 1 at.%.

### 4. Embrittlement of Grain Boundaries from Solute Segregation

Solute segregation can induce bonding changes at GBs, thereby encouraging (or discouraging) intergranular fracture. Rice-Wang thermodynamic analysis can provide insights into bond weakening or grain boundary embrittlement [22,27]. Here we conduct ab initio calculations to estimate embrittlement potency of solute segregation to grain boundary sites in the Σ5[001](210) boundary using slab calculations. We define the embrittlement potency ($\Delta E_i^{\text{emb}}$) as:

$$\Delta E_i^{\text{emb}} = \Delta E_i^{\text{GB}} - \Delta E_i^{\text{FS}} = \left(E_{\text{sol},i}^{\text{GB}} - E_{\text{pure}}^{\text{GB}}\right) - \left(E_{\text{sol},i}^{\text{FS}} - E_{\text{pure}}^{\text{FS}}\right) \tag{3}$$

where $E_{\text{sol},i}^{\text{GB}}$ and $E_{\text{sol},i}^{\text{FS}}$ denote the system energy with a solute occupying site $i$, and $E_{\text{pure}}^{\text{GB}}$ and $E_{\text{pure}}^{\text{FS}}$ are the system energy of the unalloyed GB and free surface respectively. Positive values here indicate that the solute prefers the free surface environment, and thus would induce grain boundary embrittlement due to the decrease in surface energy upon fracture relative to the unalloyed condition.

We first show a periodic summary plot in Fig. 5 for embrittlement potency and Fig 6 for the preferred fracture paths for all solute elements considered. The fracture path shown here are as defined in Fig. 1. i.e. Fracture path I indicates a solute slab with Al on site I, and path II is the opposite structure without site 1. For instance, our calculations on Al(Zn) are in agreement with Ref. [27], in that site 2 and 3 prefer to fracture with path II (no site 1 Aluminium atom adjacent to the solute on the free surface).

Fig. 6 also implies that it is important to consider multiple fracture paths in GBs, although with the sheer number of possible fracture paths, this may not be tractable for DFT calculations. With the advent of machine learning in materials science, this may become possible in the near future. To estimate the important of fracture path effects, we show in Fig. 7 the difference between embrittlement potency of both fracture paths. For instance, embrittlement potency of Al(Sc, Zr and W) at site 2 can be more than 1 eV different. This stresses the importance of adopting a statistical perspective in grain boundary embrittlement problems [6]. Due to the sheer number of possible fracture paths needed in considering the statistical model, future physical or machine learning models that aid in determining or reducing preferential fracture paths may help modeling intergranular fracture at multiple length scales.

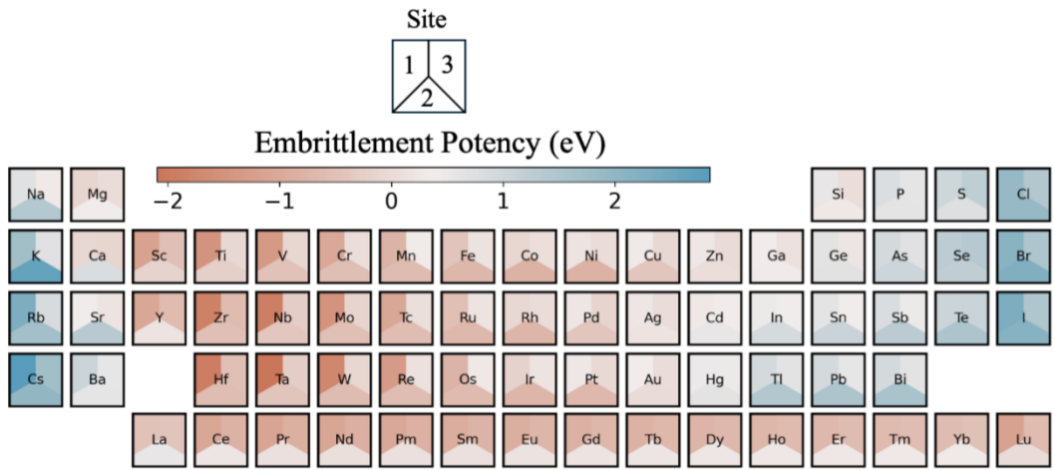

**Fig. 4** Periodic table plot for embrittlement potency of all three sites. For site 2 and 3, the most embrittling (positive) values are presented here (see supplemental Table S1 for numerical details, and Fig.5 and 6 for the fracture path preferences).

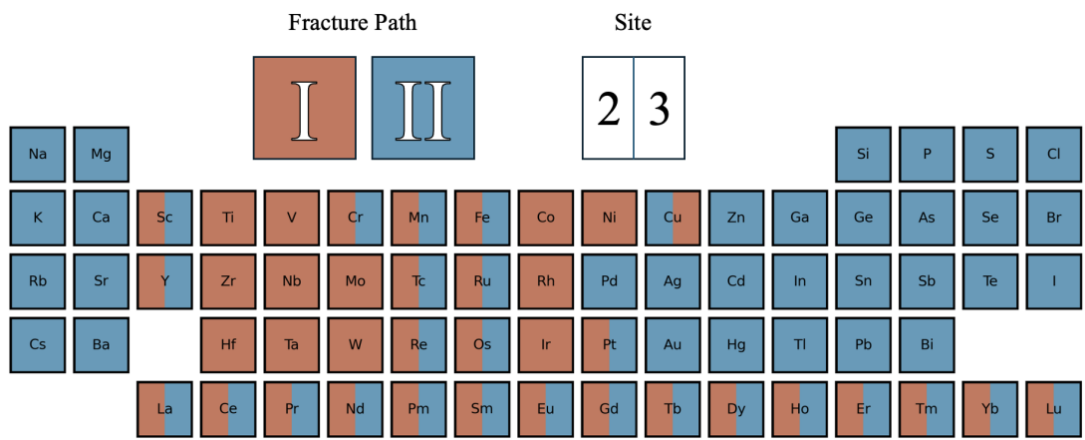

**Fig. 5** Preferred fracture path of site 2 (left) and 3 (right). The red color indicates fracture path 1 (or w/ site 1 on the solute free surface slab) and blue shade indicates the second path (w/o site 1 next to the solute free surface slab).

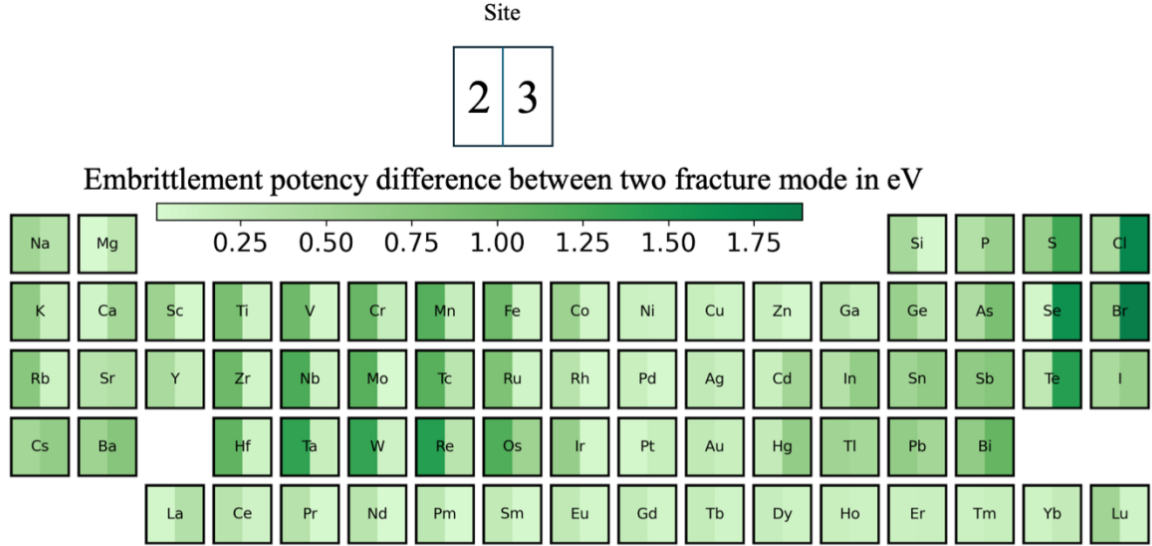

**Fig. 6** Embrittlement potency differences between two fracture paths for site 2 and 3 segregated slabs.

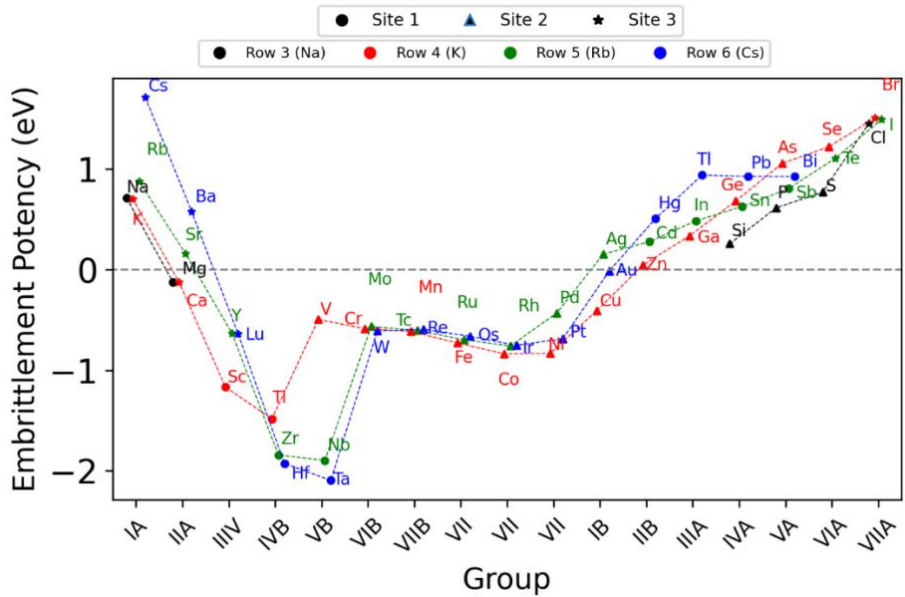

**Fig. 7** Embrittlement potencies of all the solute elements calculated. The symbols indicate preferential site for solute segregation, which the value of embrittlement potency is taken from.

## 5. Statistical and Solubility Perspectives of Grain Boundary Embrittlement

To average the embrittlement potency an individual boundary [6], we need the site occupation probabilities which can be estimated via Eq. (2). The averaged $\Delta E^{\mathrm{emb}}$ of a boundary can then be computed as:

$$\langle \Delta E^{\text{emb}} \rangle = \sum_i F_i X_i^{\text{GB}} \Delta E_i^{\text{emb}} \quad (4)$$

with $F_i$ as the site fraction for site type '$i$'. For each GB site, we choose the most embrittling fracture path, on the grounds that nature would find the most easily fractured path and thus is an upper-bound selector for $\Delta E^{\text{emb}}$. Eq. (4) weights each site according to its solute occupation probability (note that site 2 and 3 are more degenerated than site 1), and thus accounts for a true expected segregation state rather than an artificial one by accounting for unoccupied sites that fracture at the intrinsic cohesion energy of the boundary itself. Fig. 8a is a bar chart calculated with $X^{\text{tot}}$ of 1 at. % and $T = 475$ °C, which is a typical heat treatment temperature at which many Al alloys might be reasonably equilibrated. Here again we see examples where there are transition metal solute elements that enhance the cohesion of the GB (i.e. V, Cr and Nb for example), they do not segregate well and thus do not significantly contribute to the enhancement of GB cohesion.

What is more concerning is that several strengthening and segregating elements here are, in fact, not very soluble in an FCC Al matrix, so the trend in Fig. 8a is not reflective of the true potential they have for improving GB cohesion. As a means of including this consideration, we can apply a thermodynamic criterion to account for their solubility limit in cases where it is below the nominal bulk composition of 1 at%:

$$X^C = \min\ (1\ at.\%, \text{Solubility from TCAL9}) \quad (5)$$

We obtain the solubility at $T = 475$ °C from Thermo-Calc [75] TCAL9 database [76], and plot a ranked bar plot in Fig. 8b, limiting the analysis only to systems with available thermodynamic data. This bar chart shows the net effect of both segregation and embrittlement, while the presentation in Fig.9 juxtaposes those two contributions explicitly against each other to make clear which physics dominate the net effect. The strongest candidates for GB cohesivity are Y and Ni due to their strong tendency of segregation (cf. Fig. 2 and 3) despite their relatively low solubility, combined with their clear strengthening effect at the GB. However, several promising candidates such as Sc and Zr, nominally identified as potential cohesion enhancers, have lower segregation energies and thus low GB concentrations. For instance, Al(Sc) and Al(Zr) show drops in the embrittlement products from −0.21 and −0.18 eV to −0.012 and −0.004 respectively at $X^{\text{tot}}$ of 1 at.% and $T = 475$ °C due to their solubility limit computed from a thermodynamic database (0.034 and 0.014 at.% respectively). The solubility limits reflect in the GB concentration of only 1.7 (Sc) and 0.3 at.% (Zr). Similar problems may arise for interstitial solutes such as Al(B) as well [28]. Another example here is Hf; while showing GB cohesion enhancement, low tendency of segregation may limit its practicality for Al GBs (cf. Fig. 9). Lastly, Sn and Pb, which are elements with low melting point and cohesive energy, show strong tendency of GB embrittlement.

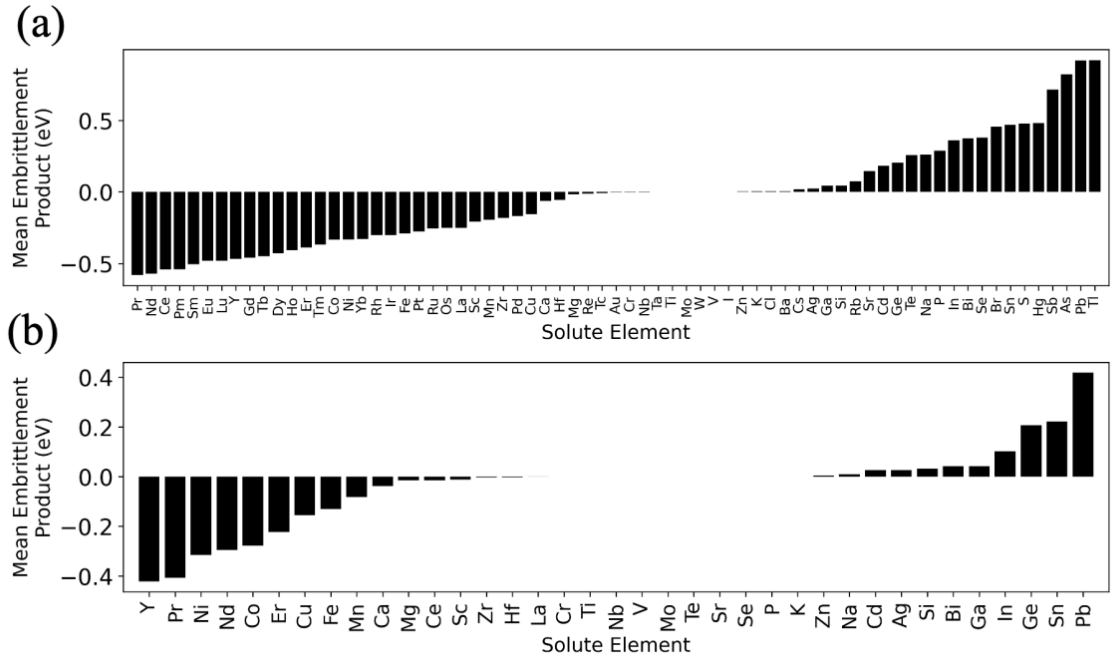

**Fig. 8** Bar chart for (a) all solute elements assuming bulk concentration of 1 at.% and (b) with solubility constraints from TCAL9 database (1 at.% or solubility at 475 °C, whichever is lower).

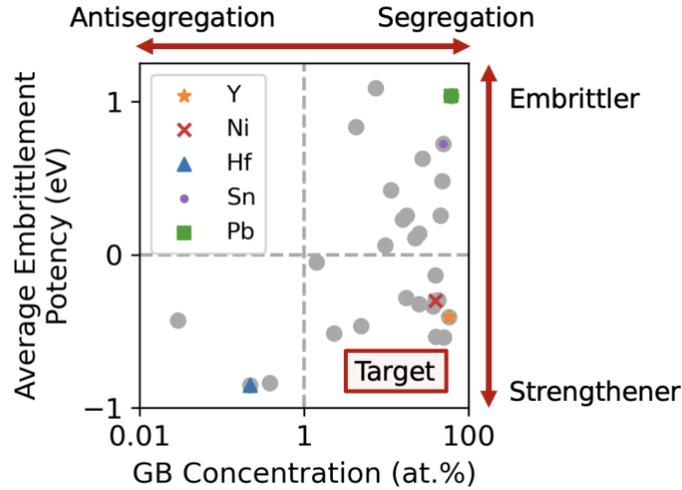

**Fig. 9** Grain boundary concentration-embrittlement potency plot for all alloys. The grain boundary concentrations are calculated at $X^C$ = 1 at.% (shown via a vertical dashed line) and $T$ = 748 K (475 °C).

## 6. On the Variation of Density Functional Theory Methods

The present systematic all-electron computations provide an opportunity to benchmark both segregation and embrittlement using the more common PAW method with Vienna Ab Initio Software Package (VASP) [77–81] excluding the lanthanides. We show parity plots of segregation in Fig. 10a and embrittlement in Fig. 10b. These data show excellent agreement, except for a few systems such as As and Rb. Such systems are rarely considered for practical metallurgy, and even so, the errors, for instance at the most segregating sites, show an embrittlement potency of 0.84 eV for site 1 Al(As) and 1.15 eV for site 3 Al(Rb) (vis-à-vis the FLAPW potency of 0.94 and 0.87 eV respectively); both solute species are strong GB embrittlers with either method. The benchmarking exercise thus suggests that PAW method is likely accurate enough for most cases. Variations seen amongst first-principles data [19] may thus stem from other considerations: varying pseudopotentials, k-point meshes, planewave cutoff energies, supercell sizes and relaxation methods [82]. For instance, we show here a calculation of Al(Sc) with Hubbard $U$ density functional theory (DFT+U) via fleur [83–85] (following the $U$ value of 4.2 eV used in Ref. [34]) in Table 2 and Fig. 11. While the band structure beyond the fermi energy changes noticeably, we found that the segregation energy and embrittlement potencies do not significantly change. The contrast between the magnitude of segregation energy in Ref. [34] (which is of order eV) and Ref. [18] (which shows only an order of magnitude less for Σ5[001](210)) could stem from a different bulk reference state, or different relaxation and other parameters used in the calculations.

These results suggests that comparative efforts with consistent methodology are needed to provide accurate relative segregation and embrittlement tendencies of substitutional alloys across the same base metal system. Note that while we show the results for several elements that may pose limited accuracy due to localization of electrons such as lanthanides, the results are subjected to further investigation. Despite the better accuracy of methodologies such as hybrid functionals [40,86–88] that aim to alleviate the problem, their computational cost is still too intensive for high throughput calculations with the GB slabs used in this work.

**Table 2** Segregation energy and embrittlement potency of Al(Sc) with (bolded) and without DFT+U = 4.2 eV (in bracket) calculated with the FLAPW method.

| Site | $\Delta E^{seg}$ (eV) | $\Delta E^{coh}$ (eV) |
| --- | --- | --- |
| 1 | **−0.41** (-0.29) | **−1.16** (-1.17) |
| 2-1 | **0.10** (0.12) | **−0.03** (-0.17) |
| 2-2 | | **−0.61** (-0.73) |
| 3-1 | **−0.41** (-0.29) | **−0.62** (-0.57) |
| 3-2 | | **−0.50** (-0.53) |

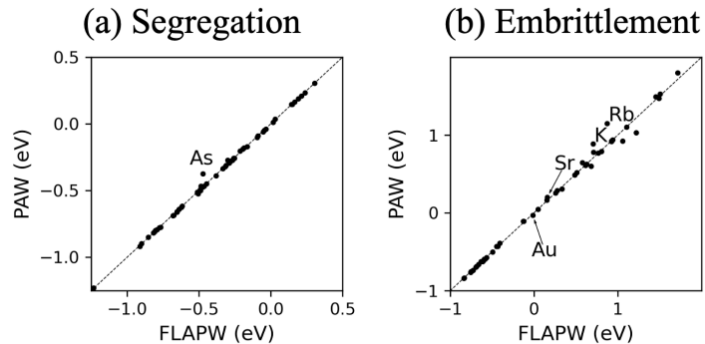

**Fig. 10** (a) Grain boundary concentration and (b) embrittlement potency calculated from FLAPW and PAW method described here. The embrittlement potencies in (b) are taken from the most segregating site. (see supplemental Table S2 for the numerical data from the PAW method).

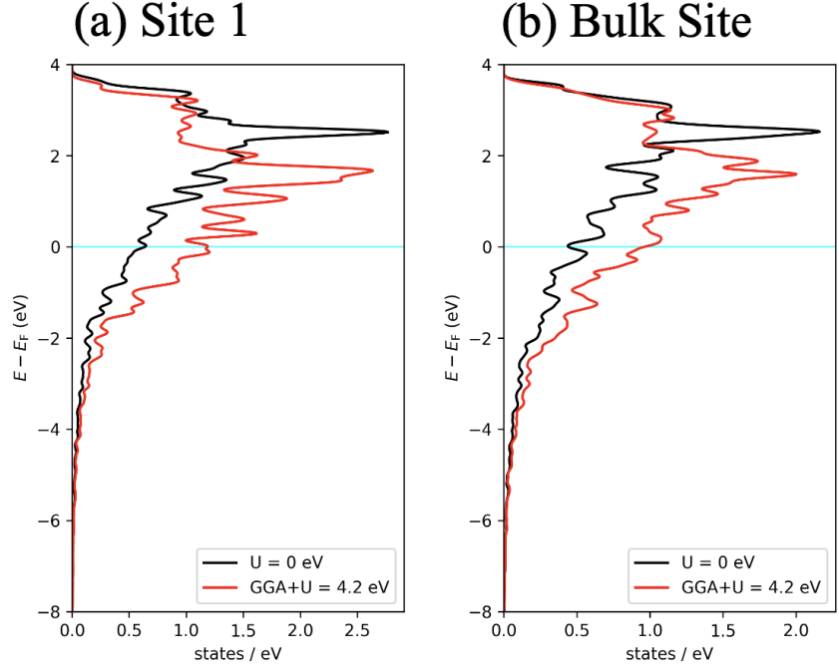

**Fig. 11** Density of state for Sc d-band at (a) site 1 and (b) the bulk site, both with and without Hubbard $U$ density functional theory with $U = 4.2$ eV.

7. **Conclusions**

In summary, we have computed grain boundary segregation and embrittlement potencies of Al Σ5[001](210) GB for 69 substitutional binary solute elements, with an all-electron FLAPW DFT methodology. We show that, first, it is important to consider multiple possible fracture paths near doped solute sites, as several transition metal systems show preferences for fracture adjacent a GB-centered Al atom instead of being directly adjacent to the free surface. Not only does the fracture path influence the alloy design decisions, but the solubility of solute elements also plays an important role; for multiple solute species that prefer to segregate and enhance or embrittle the GB, it is found that their solubility is often low in Al matrix and thus limit their influence over GB cohesion (for instance, Al(Sc) and Al(Zr) show drops in the cohesion enhancement by an order of magnitude at 1 at.% loading). In addition to the FLAPW results, PAW calculations also suggest that for the cases of Al alloys, PAW pseudopotentials provide great enough accuracy for the case of GB thermodynamic problems. By extension, variations in previously reported calculation results may stem from variation in methodologies and parameters in DFT implementation. We hope the present a set of GB data will provide an enduring reference library for future Al-based alloy design with resistance to intergranular fracture and stress corrosion cracking.

8. **Acknowledgements**

The work is supported by Office of Naval Research (ONR) under the grant N000142312004. The authors acknowledge a support from Constellium. The authors would like to acknowledge MIT ORCD, MIT Engaging and MIT Supercloud [89] for the HPC resources used in this work. N. Tuchinda acknowledges fruitful discussions with O.Y. Kontsevoi, C. Li and M. Tzini.

## 9. References

[1] J. Bruley, V.J. Keast, D.B. Williams, An EELS study of segregation-induced grain-boundary embrittlement of copper, Acta Mater. 47 (1999) 4009–4017. https://doi.org/10.1016/S1359-6454(99)00261-X.

[2] R. Wu, A.J. Freeman, G.B. Olson, Nature of phosphorus embrittlement of the FeΣ3[11$\bar{0}$](111) grain boundary, Phys. Rev. B 50 (1994) 75–81. https://doi.org/10.1103/PhysRevB.50.75.

[3] M.A. Gibson, C.A. Schuh, Segregation-induced changes in grain boundary cohesion and embrittlement in binary alloys, Acta Mater. 95 (2015) 145–155. https://doi.org/10.1016/j.actamat.2015.05.004.

[4] P. Lejček, Grain Boundary Segregation in Metals, Springer Berlin Heidelberg, Berlin, Heidelberg, 2010. https://doi.org/10.1007/978-3-642-12505-8.

[5] M.E. Eberhart, D.D. Vvedensky, Localized Grain-Boundary Electronic States and Intergranular Fracture, Phys. Rev. Lett. 58 (1987) 61–64. https://doi.org/10.1103/PhysRevLett.58.61.

[6] M.E. Fernandez, R. Dingreville, D.E. Spearot, Statistical perspective on embrittling potency for intergranular fracture, Phys. Rev. Mater. 6 (2022) 083602. https://doi.org/10.1103/PhysRevMaterials.6.083602.

[7] M.E. Fernandez, R. Dingreville, D.L. Medlin, D.E. Spearot, The Effect of Grain Boundary Facet Junctions on Segregation and Embrittlement, Acta Mater. 269 (2024) 119805. https://doi.org/10.1016/j.actamat.2024.119805.

[8] J. Weissmüller, Alloy effects in nanostructures, Proc. First Int. Conf. Nanostructured Mater. 3 (1993) 261–272. https://doi.org/10.1016/0965-9773(93)90088-S.

[9] R. Kirchheim, Reducing grain boundary, dislocation line and vacancy formation energies by solute segregation. I. Theoretical background, Acta Mater. 55 (2007) 5129–5138. https://doi.org/10.1016/j.actamat.2007.05.047.

[10] A.R. Kalidindi, T. Chookajorn, C.A. Schuh, Nanocrystalline Materials at Equilibrium: A Thermodynamic Review, JOM 67 (2015) 2834–2843. https://doi.org/10.1007/s11837-015-1636-9.

[11] T.P. Matson, C.A. Schuh, Phase and defect diagrams based on spectral grain boundary segregation: A regular solution approach, Acta Mater. 265 (2024) 119584. https://doi.org/10.1016/j.actamat.2023.119584.

[12] F. Abdeljawad, P. Lu, N. Argibay, B.G. Clark, B.L. Boyce, S.M. Foiles, Grain boundary segregation in immiscible nanocrystalline alloys, Acta Mater. 126 (2017) 528–539. https://doi.org/10.1016/j.actamat.2016.12.036.

[13] P. Lu, F. Abdeljawad, M. Rodriguez, M. Chandross, D.P. Adams, B.L. Boyce, B.G. Clark, N. Argibay, On the thermal stability and grain boundary segregation in nanocrystalline PtAu alloys, Materialia 6 (2019) 100298. https://doi.org/10.1016/j.mtla.2019.100298.

[14] A.R. Miedema, Surface Segregation in Alloys of Transition Metals, Int. J. Mater. Res. 69 (1978) 455–461. https://doi.org/doi:10.1515/ijmr-1978-690706.

[15] H.A. Murdoch, C.A. Schuh, Estimation of grain boundary segregation enthalpy and its role in stable nanocrystalline alloy design, J. Mater. Res. 28 (2013) 2154–2163. https://doi.org/10.1557/jmr.2013.211.

[16] L. Huber, B. Grabowski, M. Militzer, J. Neugebauer, J. Rottler, Ab initio modelling of solute segregation energies to a general grain boundary, Acta Mater. 132 (2017) 138–148. https://doi.org/10.1016/j.actamat.2017.04.024.

[17] L. Huber, R. Hadian, B. Grabowski, J. Neugebauer, A machine learning approach to model solute grain boundary segregation, Npj Comput. Mater. 4 (2018) 64. https://doi.org/10.1038/s41524-018-0122-7.

[18] R. Mahjoub, K.J. Laws, N. Stanford, M. Ferry, General trends between solute segregation tendency and grain boundary character in aluminum - An ab inito study, Acta Mater. 158 (2018) 257–268. https://doi.org/10.1016/j.actamat.2018.07.069.

[19] M.A. Gibson, C.A. Schuh, A survey of ab-initio calculations shows that segregation-induced grain boundary embrittlement is predicted by bond-breaking arguments, Scr. Mater. 113 (2016) 55–58. https://doi.org/10.1016/j.scriptamat.2015.09.041.

[20] H. Erhart, H.J. Grabke, Equilibrium segregation of phosphorus at grain boundaries of Fe–P, Fe–C–P, Fe–Cr–P, and Fe–Cr–C–P alloys, Met. Sci. 15 (1981) 401–408. https://doi.org/10.1179/030634581790426877.

[21] D. Raabe, M. Herbig, S. Sandlöbes, Y. Li, D. Tytko, M. Kuzmina, D. Ponge, P.-P. Choi, Grain boundary segregation engineering in metallic alloys: A pathway to the design of interfaces, Curr. Opin. Solid State Mater. Sci. 18 (2014) 253–261. https://doi.org/10.1016/j.cossms.2014.06.002.

[22] J.R. Rice, J.-S. Wang, Embrittlement of interfaces by solute segregation, Poceedings Symp. Interfacial Phenom. Compos. Process. Charact. Mech. Prop. 107 (1989) 23–40. https://doi.org/10.1016/0921-5093(89)90372-9.

[23] H. Zhao, P. Chakraborty, D. Ponge, T. Hickel, B. Sun, C.-H. Wu, B. Gault, D. Raabe, Hydrogen trapping and embrittlement in high-strength Al alloys, Nature 602 (2022) 437–441. https://doi.org/10.1038/s41586-021-04343-z.

[24] H.L. Mai, X.-Y. Cui, D. Scheiber, L. Romaner, S.P. Ringer, The segregation of transition metals to iron grain boundaries and their effects on cohesion, Acta Mater. 231 (2022) 117902. https://doi.org/10.1016/j.actamat.2022.117902.

[25] S. Zhang, O.Y. Kontsevoi, A.J. Freeman, G.B. Olson, Cohesion enhancing effect of magnesium in aluminum grain boundary: A first-principles determination, Appl. Phys. Lett. 100 (2012) 231904. https://doi.org/10.1063/1.4725512.

[26] S. Zhang, O.Y. Kontsevoi, A.J. Freeman, G.B. Olson, Sodium-induced embrittlement of an aluminum grain boundary, Phys. Rev. B 82 (2010) 224107. https://doi.org/10.1103/PhysRevB.82.224107.

[27] S. Zhang, O.Y. Kontsevoi, A.J. Freeman, G.B. Olson, First principles investigation of zinc-induced embrittlement in an aluminum grain boundary, Acta Mater. 59 (2011) 6155–6167. https://doi.org/10.1016/j.actamat.2011.06.028.

[28] S. Zhang, O.Y. Kontsevoi, A.J. Freeman, G.B. Olson, First-principles determination of the effect of boron on aluminum grain boundary cohesion, Phys. Rev. B 84 (2011) 134104. https://doi.org/10.1103/PhysRevB.84.134104.

[29] R.G. Song, W. Dietzel, B.J. Zhang, W.J. Liu, M.K. Tseng, A. Atrens, Stress corrosion cracking and hydrogen embrittlement of an Al–Zn–Mg–Cu alloy, Acta Mater. 52 (2004) 4727–4743. https://doi.org/10.1016/j.actamat.2004.06.023.

[30] S. Zhang, O.Y. Kontsevoi, A.J. Freeman, G.B. Olson, Aluminum grain boundary decohesion by dense sodium segregation, Phys. Rev. B 85 (2012) 214109. https://doi.org/10.1103/PhysRevB.85.214109.

[31] P. Lejček, M. Šob, V. Paidar, Interfacial segregation and grain boundary embrittlement: An overview and critical assessment of experimental data and calculated results, Prog. Mater. Sci. 87 (2017) 83–139. https://doi.org/10.1016/j.pmatsci.2016.11.001.

[32] E. Bosoni, L. Beal, M. Bercx, P. Blaha, S. Blügel, J. Bröder, M. Callsen, S. Cottenier, A. Degomme, V. Dikan, K. Eimre, E. Flage-Larsen, M. Fornari, A. Garcia, L. Genovese, M. Giantomassi, S.P. Huber, H. Janssen, G. Kastlunger, M. Krack, G. Kresse, T.D. Kühne, K. Lejaeghere, G.K.H. Madsen, M. Marsman, N. Marzari, G. Michalicek, H. Mirhosseini, T.M.A. Müller, G. Petretto, C.J. Pickard, S. Poncé, G.-M. Rignanese, O. Rubel, T. Ruh, M. Sluydts, D.E.P. Vanpoucke, S. Vijay, M. Wolloch, D. Wortmann, A.V. Yakutovich, J. Yu, A. Zadoks, B. Zhu, G. Pizzi, How to verify the precision of density-functional-theory implementations via reproducible and universal workflows, Nat. Rev. Phys. 6 (2023) 45–58. https://doi.org/10.1038/s42254-023-00655-3.

[33] Z. Xiao, J. Hu, Y. Liu, F. Dong, Y. Huang, Segregation of Sc and its effects on the strength of Al Σ5 (210) [100] symmetrical tilt grain boundary, Mater. Sci. Eng. A 756 (2019) 389–395. https://doi.org/10.1016/j.msea.2019.04.070.

[34] Z. Xiao, J. Hu, Y. Liu, F. Dong, Y. Huang, Co-segregation behavior of Sc and Zr solutes and their effect on the Al Σ5 (210) [110] symmetrical tilt grain boundary: a first-principles study, Phys. Chem. Chem. Phys. 21 (2019) 19437–19446. https://doi.org/10.1039/C9CP03002F.

[35] O.K. Andersen, Linear methods in band theory, Phys. Rev. B 12 (1975) 3060–3083. https://doi.org/10.1103/PhysRevB.12.3060.

[36] E. Wimmer, H. Krakauer, M. Weinert, A.J. Freeman, Full-potential self-consistent linearized-augmented-plane-wave method for calculating the electronic structure of molecules and surfaces: ${\mathrm{O}}_{2}$ molecule, Phys. Rev. B 24 (1981) 864–875. https://doi.org/10.1103/PhysRevB.24.864.

[37] M. Weinert, E. Wimmer, A.J. Freeman, Total-energy all-electron density functional method for bulk solids and surfaces, Phys. Rev. B 26 (1982) 4571–4578. https://doi.org/10.1103/PhysRevB.26.4571.

[38] M. Weinert, G. Schneider, R. Podloucky, J. Redinger, FLAPW: applications and implementations, J. Phys. Condens. Matter 21 (2009) 084201. https://doi.org/10.1088/0953-8984/21/8/084201.

[39] H. Krakauer, M. Posternak, A.J. Freeman, Linearized augmented plane-wave method for the electronic band structure of thin films, Phys. Rev. B 19 (1979) 1706–1719. https://doi.org/10.1103/PhysRevB.19.1706.

[40] D. Singh, Ground-state properties of lanthanum: Treatment of extended-core states, Phys. Rev. B 43 (1991) 6388–6392. https://doi.org/10.1103/PhysRevB.43.6388.

[41] R. Yu, D. Singh, H. Krakauer, All-electron and pseudopotential force calculations using the linearized-augmented-plane-wave method, Phys. Rev. B 43 (1991) 6411–6422. https://doi.org/10.1103/PhysRevB.43.6411.

[42] D.A. Klüppelberg, M. Betzinger, S. Blügel, Atomic force calculations within the all-electron FLAPW method: Treatment of core states and discontinuities at the muffin-tin sphere boundary, Phys. Rev. B 91 (2015) 035105. https://doi.org/10.1103/PhysRevB.91.035105.

[43] E. Şaşıoğlu, C. Friedrich, S. Blügel, Effective Coulomb interaction in transition metals from constrained random-phase approximation, Phys. Rev. B 83 (2011) 121101. https://doi.org/10.1103/PhysRevB.83.121101.

[44] D. Wortmann, G. Michalicek, N. Baadji, M. Betzinger, G. Bihlmayer, J. Bröder, T. Burnus, J. Enkovaara, F. Freimuth, C. Friedrich, C.-R. Gerhorst, S. Granberg Cauchi, U. Grytsiuk, A. Hanke, J.-P. Hanke, M. Heide, S. Heinze, R. Hilgers, H. Janssen, D.A. Klüppelberg, R. Kovacik, P. Kurz, M. Lezaic, G.K.H. Madsen, Y. Mokrousov, A. Neukirchen, M. Redies, S. Rost, M. Schlipf, A. Schindlmayr, M. Winkelmann, S. Blügel, FLEUR, (2023). https://doi.org/10.5281/zenodo.7891361.

[45] A. Freeman, R. Wu, Electronic structure theory of surface, interface and thin-film magnetism, J. Magn. Magn. Mater. 100 (1991) 497–514.

[46] J.P. Perdew, K. Burke, M. Ernzerhof, Generalized Gradient Approximation Made Simple, Phys. Rev. Lett. 77 (1996) 3865–3868. https://doi.org/10.1103/PhysRevLett.77.3865.

[47] A. Stukowski, Visualization and analysis of atomistic simulation data with OVITO–the Open Visualization Tool, Model. Simul. Mater. Sci. Eng. 18 (2009) 015012. https://doi.org/10.1088/0965-0393/18/1/015012.

[48] G. vanRossum, Python reference manual, Dep. Comput. Sci. CS (1995).

[49] M.L. Waskom, Seaborn: statistical data visualization, J. Open Source Softw. 6 (2021) 3021.

[50] J.D. Hunter, Matplotlib: A 2D graphics environment, Comput. Sci. Eng. 9 (2007) 90–95. https://doi.org/10.1109/MCSE.2007.55.

[51] J. Riebesell, H. (Daniel) Yang, R. Goodall, S.G. Baird, H. Zheng, J. George, janosh/pymatviz: v0.11.0, (2024). https://doi.org/10.5281/zenodo.13624216.

[52] A.H. Larsen, J.J. Mortensen, J. Blomqvist, I.E. Castelli, R. Christensen, M. Dułak, J. Friis, M.N. Groves, B. Hammer, C. Hargus, The atomic simulation environment—a Python library for working with atoms, J. Phys. Condens. Matter 29 (2017) 273002.

[53] I. Flyamer, Z. Xue, Colin, A. Li, R. Neff, V. Vazquez, S. Dicks, O. Gustafsson, N. Morshed, J.L. Espinoza, JasonMendoza2008, mski_iksm, N. Vaulin, Sandro, S. Edelbrock, scaine1, 136s, O. Lee, Phlya/adjustText: 1.3.0, (2024). https://doi.org/10.5281/zenodo.14019059.

[54] C.R. Harris, K.J. Millman, S.J. van der Walt, R. Gommers, P. Virtanen, D. Cournapeau, E. Wieser, J. Taylor, S. Berg, N.J. Smith, R. Kern, M. Picus, S. Hoyer, M.H. van Kerkwijk, M. Brett, A. Haldane, J.F. del Río, M. Wiebe, P. Peterson, P. Gérard-Marchant, K. Sheppard, T. Reddy, W. Weckesser, H. Abbasi, C. Gohlke, T.E. Oliphant, Array programming with NumPy, Nature 585 (2020) 357–362. https://doi.org/10.1038/s41586-020-2649-2.

[55] P. Virtanen, R. Gommers, T.E. Oliphant, M. Haberland, T. Reddy, D. Cournapeau, E. Burovski, P. Peterson, W. Weckesser, J. Bright, S.J. van der Walt, M. Brett, J. Wilson, K.J. Millman, N. Mayorov, A.R.J. Nelson, E. Jones, R. Kern, E. Larson, C.J. Carey, İ. Polat, Y. Feng, E.W. Moore, J. VanderPlas, D. Laxalde, J. Perktold, R. Cimrman, I. Henriksen, E.A. Quintero, C.R. Harris, A.M. Archibald, A.H. Ribeiro, F. Pedregosa, P. van Mulbregt, A. Vijaykumar, A.P. Bardelli, A. Rothberg, A. Hilboll, A. Kloeckner, A. Scopatz, A. Lee, A. Rokem, C.N. Woods, C. Fulton, C. Masson, C. Häggström, C. Fitzgerald, D.A. Nicholson, D.R. Hagen, D.V. Pasechnik, E. Olivetti, E. Martin, E. Wieser, F. Silva, F. Lenders, F. Wilhelm, G. Young, G.A. Price, G.-L. Ingold, G.E. Allen, G.R. Lee, H. Audren, I. Probst, J.P. Dietrich, J. Silterra, J.T. Webber, J. Slavič, J. Nothman, J. Buchner, J. Kulick, J.L. Schönberger, J.V. de Miranda Cardoso, J. Reimer, J. Harrington, J.L.C. Rodríguez, J. Nunez-Iglesias, J. Kuczynski, K. Tritz, M. Thoma, M. Newville, M. Kümmerer, M. Bolingbroke, M. Tartre, M. Pak, N.J. Smith, N. Nowaczyk, N. Shebanov, O. Pavlyk, P.A. Brodtkorb, P. Lee, R.T. McGibbon, R. Feldbauer, S. Lewis, S. Tygier, S. Sievert, S. Vigna, S. Peterson, S. More, T. Pudlik, T. Oshima, T.J. Pingel, T.P. Robitaille, T. Spura, T.R. Jones, T. Cera, T. Leslie, T. Zito, T. Krauss, U. Upadhyay, Y.O. Halchenko, Y. Vázquez-Baeza, SciPy 1.0 Contributors, SciPy 1.0: fundamental algorithms for scientific computing in Python, Nat. Methods 17 (2020) 261–272. https://doi.org/10.1038/s41592-019-0686-2.

[56] Bokeh Development Team, Bokeh: Python library for interactive visualization, 2018. https://bokeh.pydata.org/en/latest/.

[57] K.M. Thyng, C.A. Greene, R.D. Hetland, H.M. Zimmerle, S.F. DiMarco, True Colors of Oceanography, Oceanography 29 (2016) 9–13.

[58] T. Kluyver, B. Ragan-Kelley, F. Pérez, B. Granger, M. Bussonnier, J. Frederic, K. Kelley, J. Hamrick, J. Grout, S. Corlay, P. Ivanov, D. Avila, S. Abdalla, C. Willing, Jupyter Notebooks – a publishing format for reproducible computational workflows, in: F. Loizides, B. Schmidt (Eds.), Position. Power Acad. Publ. Play. Agents Agendas, IOS Press, 2016: pp. 87–90.

[59] G.C. Hasson, C. Goux, Interfacial energies of tilt boundaries in aluminium. Experimental and theoretical determination, Scr. Metall. 5 (1971) 889–894. https://doi.org/10.1016/0036-9748(71)90064-0.

[60] D. Zhao, O.M. Løvvik, K. Marthinsen, Y. Li, Segregation of Mg, Cu and their effects on the strength of Al Σ5 (210)[001] symmetrical tilt grain boundary, Acta Mater. 145 (2018) 235–246. https://doi.org/10.1016/j.actamat.2017.12.023.

[61] R. Hadian, B. Grabowski, J. Neugebauer, GB code: A grain boundary generation code, J. Open Source Softw. 3 (2018). https://doi.org/10.21105/joss.00900.

[62] D. McLean, Grain boundaries in metals, Clarendon Press, Oxford, 1957.

[63] C. White, D. Stein, Sulfur segregation to grain boundaries in Ni 3 Al and Ni 3 (AI, Ti) alloys, Metall. Trans. A 9 (1978) 13–22.

[64] M. Wagih, Y. Naunheim, T. Lei, C.A. Schuh, Grain boundary segregation predicted by quantum-accurate segregation spectra but not by classical models, Acta Mater. 266 (2024) 119674. https://doi.org/10.1016/j.actamat.2024.119674.

[65] M. de Hass, J.Th.M. De Hosson, Grain boundary segregation and precipitation in aluminium alloys, Scr. Mater. 44 (2001) 281–286. https://doi.org/10.1016/S1359-6462(00)00577-7.

[66] A. Garner, R. Euesden, Y. Yao, Y. Aboura, H. Zhao, J. Donoghue, M. Curioni, B. Gault, P. Shanthraj, Z. Barrett, C. Engel, T.L. Burnett, P.B. Prangnell, Multiscale analysis of grain boundary microstructure in high strength 7xxx Al alloys, Acta Mater. 202 (2021) 190–210. https://doi.org/10.1016/j.actamat.2020.10.021.

[67] W. Xing, A.R. Kalidindi, D. Amram, C.A. Schuh, Solute interaction effects on grain boundary segregation in ternary alloys, Acta Mater. 161 (2018) 285–294. https://doi.org/10.1016/j.actamat.2018.09.005.

[68] M. Wagih, Y. Naunheim, T. Lei, C.A. Schuh, Designing for Cooperative Grain Boundary Segregation in Multicomponent Alloys, (n.d.).

[69] M. Guttmann, Equilibrium segregation in a ternary solution: A model for temper embrittlement, Surf. Sci. 53 (1975) 213–227. https://doi.org/10.1016/0039-6028(75)90125-9.

[70] P. Lejček, Effect of ternary solute interaction on interfacial segregation and grain boundary embrittlement, J. Mater. Sci. 48 (2013) 4965–4972.

[71] J.W. Cahn, Thermodynamics of solid and fluid surfaces, Sel. Works John W Cahn (1998) 379–399.

[72] W.C. Carter, W.C. Johnson, The selected works of John W. Cahn, John Wiley & Sons, 2013.

[73] A. Reichmann, N. Tuchinda, C. Dösinger, D. Scheiber, V. Razumovskiy, O. Peil, T.P. Matson, C.A. Schuh, L. Romaner, Grain boundary segregation for the Fe-P system: Insights from atomistic modeling and Bayesian inference, Acta Mater. (2024) 120215. https://doi.org/10.1016/j.actamat.2024.120215.

[74] M. Wagih, C.A. Schuh, Spectrum of grain boundary segregation energies in a polycrystal, Acta Mater. 181 (2019) 228–237. https://doi.org/10.1016/j.actamat.2019.09.034.

[75] J.-O. Andersson, T. Helander, L. Höglund, P. Shi, B. Sundman, Thermo-Calc & DICTRA, computational tools for materials science, Calphad 26 (2002) 273–312. https://doi.org/10.1016/S0364-5916(02)00037-8.

[76] Thermo-Calc Software TCAL Aluminium/Al-alloys Database version 9, (2024). https://thermocalc.com/products/databases/aluminum-based-alloys/.

[77] G. Kresse, J. Hafner, Ab initio molecular dynamics for liquid metals, Phys. Rev. B 47 (1993) 558–561. https://doi.org/10.1103/PhysRevB.47.558.

[78] G. Kresse, J. Furthmüller, Efficiency of ab-initio total energy calculations for metals and semiconductors using a plane-wave basis set, Comput. Mater. Sci. 6 (1996) 15–50. https://doi.org/10.1016/0927-0256(96)00008-0.

[79] G. Kresse, J. Furthmüller, Efficient iterative schemes for ab initio total-energy calculations using a plane-wave basis set, Phys. Rev. B 54 (1996) 11169–11186. https://doi.org/10.1103/PhysRevB.54.11169.

[80] G. Kresse, D. Joubert, From ultrasoft pseudopotentials to the projector augmented-wave method, Phys. Rev. B 59 (1999) 1758–1775. https://doi.org/10.1103/PhysRevB.59.1758.

[81] G. Kresse, J. Furthmüller, J. Hafner, Theory of the crystal structures of selenium and tellurium: the effect of generalized-gradient corrections to the local-density approximation, Phys. Rev. B 50 (1994) 13181.

[82] K. Lejaeghere, G. Bihlmayer, T. Björkman, P. Blaha, S. Blügel, V. Blum, D. Caliste, I.E. Castelli, S.J. Clark, A. Dal Corso, S. De Gironcoli, T. Deutsch, J.K. Dewhurst, I. Di Marco, C. Draxl, M. Dułak, O. Eriksson, J.A. Flores-Livas, K.F. Garrity, L. Genovese, P. Giannozzi, M. Giantomassi, S. Goedecker, X. Gonze, O. Grånäs, E.K.U. Gross, A. Gulans, F. Gygi, D.R. Hamann, P.J. Hasnip, N.A.W. Holzwarth, D. Iușan, D.B. Jochym, F. Jollet, D. Jones, G. Kresse, K. Koepernik, E. Küçükbenli, Y.O. Kvashnin, I.L.M. Locht, S. Lubeck, M. Marsman, N. Marzari, U. Nitzsche, L. Nordström, T. Ozaki, L. Paulatto, C.J. Pickard, W. Poelmans, M.I.J. Probert, K. Refson, M. Richter, G.-M. Rignanese, S. Saha, M. Scheffler, M. Schlipf, K. Schwarz, S. Sharma, F. Tavazza, P. Thunström, A. Tkatchenko, M. Torrent, D. Vanderbilt, M.J. Van Setten, V. Van Speybroeck, J.M. Wills, J.R. Yates, G.-X. Zhang, S. Cottenier, Reproducibility in density functional theory calculations of solids, Science 351 (2016) aad3000. https://doi.org/10.1126/science.aad3000.

[83] A.I. Liechtenstein, V.I. Anisimov, J. Zaanen, Density-functional theory and strong interactions: Orbital ordering in Mott-Hubbard insulators, Phys. Rev. B 52 (1995) R5467–R5470. https://doi.org/10.1103/PhysRevB.52.R5467.

[84] B. Himmetoglu, A. Floris, S. De Gironcoli, M. Cococcioni, Hubbard-corrected DFT energy functionals: The LDA+ U description of correlated systems, Int. J. Quantum Chem. 114 (2014) 14–49.

[85] V.I. Anisimov, J. Zaanen, O.K. Andersen, Band theory and Mott insulators: Hubbard U instead of Stoner I, Phys. Rev. B 44 (1991) 943–954. https://doi.org/10.1103/PhysRevB.44.943.

[86] J.P. Perdew, M. Ernzerhof, K. Burke, Rationale for mixing exact exchange with density functional approximations, J. Chem. Phys. 105 (1996) 9982–9985. https://doi.org/10.1063/1.472933.

[87] A.D. Becke, Perspective: Fifty years of density-functional theory in chemical physics, J. Chem. Phys. 140 (2014) 18A301. https://doi.org/10.1063/1.4869598.

[88] C. Adamo, V. Barone, Toward reliable density functional methods without adjustable parameters: The PBE0 model, J. Chem. Phys. 110 (1999) 6158–6170.

[89] A. Reuther, J. Kepner, C. Byun, S. Samsi, W. Arcand, D. Bestor, B. Bergeron, V. Gadepally, M. Houle, M. Hubbell, M. Jones, A. Klein, L. Milechin, J. Mullen, A. Prout, A. Rosa, C. Yee, P. Michaleas, Interactive Supercomputing on 40,000 Cores for Machine Learning and Data Analysis, in: 2018 IEEE High Perform. Extreme Comput. Conf. HPEC, 2018: pp. 1–6. https://doi.org/10.1109/HPEC.2018.8547629.

# Supplemental Material

**Table S1** Segregation Energies and Embrittlement Potencies from FLAPW method

| Solute | $\Delta E_1^{seg}$ | $\Delta E_2^{seg}$ | $\Delta E_3^{seg}$ | $\Delta E_1^{emb}$ | $\Delta E_{2-I}^{emb}$ | $\Delta E_{3-I}^{emb}$ | $\Delta E_{2-II}^{emb}$ | $\Delta E_{3-II}^{emb}$ |
|---|---|---|---|---|---|---|---|---|
| Na | −0.653 | 0.003 | −0.637 | 0.709 | 0.915 | −0.034 | 1.439 | 0.289 |
| Mg | −0.333 | 0.147 | −0.216 | −0.127 | 0.386 | −0.253 | 0.391 | 0.017 |
| Si | 0.098 | −0.277 | −0.089 | 0.558 | −0.213 | 0.003 | 0.257 | 0.037 |
| P | −0.197 | −0.491 | −0.162 | 0.787 | 0.251 | −0.003 | 0.613 | 0.606 |
| S | −0.245 | −0.652 | −0.213 | 1.044 | 0.157 | −0.046 | 0.769 | 1.238 |
| Cl | 0.191 | 0.095 | 0.032 | 1.911 | 1.484 | −0.313 | 1.919 | 1.453 |
| K | 0.423 | 1.255 | −0.033 | 1.720 | 1.961 | 0.558 | 2.647 | 0.703 |
| Ca | −0.860 | −0.014 | −0.911 | −0.082 | 0.731 | −0.621 | 0.811 | −0.126 |
| Sc | −0.288 | 0.116 | −0.287 | −1.167 | −0.171 | −0.571 | −0.737 | −0.533 |
| Ti | 0.168 | 0.465 | 0.483 | −1.483 | −0.259 | −0.078 | −1.145 | −0.166 |
| V | 0.448 | 0.305 | 0.546 | −1.356 | −0.495 | −0.111 | −1.463 | −0.163 |
| Cr | 0.603 | 0.017 | 0.513 | −1.043 | −0.586 | −0.135 | −1.605 | 0.032 |
| Mn | 0.613 | −0.379 | 0.413 | −0.785 | −0.616 | 0.128 | −1.765 | 0.302 |
| Fe | 0.562 | −0.681 | 0.067 | −0.442 | −0.729 | 0.077 | −1.685 | 0.146 |
| Co | 0.463 | −0.854 | −0.120 | −0.117 | −0.836 | 0.054 | −1.407 | −0.015 |
| Ni | 0.362 | −0.803 | −0.093 | 0.147 | −0.831 | 0.012 | −0.958 | −0.088 |
| Cu | 0.260 | −0.487 | −0.023 | 0.292 | −0.570 | −0.077 | −0.409 | −0.157 |
| Zn | −0.014 | −0.194 | −0.078 | 0.227 | −0.098 | −0.042 | 0.046 | −0.009 |
| Ga | −0.159 | −0.213 | −0.160 | 0.375 | 0.055 | −0.044 | 0.335 | 0.123 |
| Ge | −0.234 | −0.301 | −0.258 | 0.652 | 0.085 | −0.080 | 0.681 | 0.197 |
| As | −0.440 | −0.473 | −0.449 | 0.936 | 0.646 | −0.173 | 1.055 | 0.701 |
| Se | −0.121 | −0.310 | −0.194 | 1.293 | 1.165 | −0.245 | 1.219 | 1.376 |
| Br | −0.158 | −0.132 | −0.317 | 2.112 | 1.641 | −0.384 | 2.236 | 1.507 |
| Rb | 0.658 | −0.069 | −0.187 | 2.326 | 0.956 | 0.772 | 1.702 | 0.873 |
| Sr | −0.693 | 0.170 | −0.792 | 0.420 | 1.042 | −0.211 | 1.352 | 0.157 |
| Y | −0.726 | 0.034 | −0.808 | −1.074 | 0.154 | −0.793 | −0.287 | −0.633 |
| Zr | −0.254 | 0.038 | −0.215 | −1.841 | −0.536 | −0.638 | −1.516 | −0.706 |
| Nb | 0.145 | 0.418 | 0.318 | −1.896 | −0.390 | −0.304 | −1.600 | −0.399 |
| Mo | 0.411 | 0.215 | 0.520 | −1.504 | −0.565 | −0.156 | −1.701 | −0.164 |
| Tc | 0.582 | −0.088 | 0.479 | −1.075 | −0.606 | −0.213 | −1.741 | 0.077 |
| Ru | 0.548 | −0.446 | 0.295 | −0.562 | −0.697 | 0.069 | −1.538 | 0.149 |
| Rh | 0.423 | −0.623 | 0.080 | −0.069 | −0.759 | 0.016 | −1.067 | 0.001 |
| Pd | 0.223 | −0.512 | −0.039 | 0.213 | −0.581 | −0.166 | −0.432 | −0.146 |
| Ag | −0.122 | −0.259 | −0.141 | 0.207 | −0.070 | −0.115 | 0.152 | 0.018 |
| Cd | −0.471 | −0.218 | −0.314 | 0.278 | 0.420 | −0.173 | 0.565 | 0.347 |
| In | −0.616 | −0.245 | −0.458 | 0.484 | 0.579 | −0.242 | 0.917 | 0.408 |
| Sn | −0.638 | −0.279 | −0.515 | 0.627 | 0.555 | −0.272 | 1.098 | 0.398 |
| Sb | −0.675 | −0.348 | −0.593 | 0.805 | 0.633 | −0.298 | 1.318 | 0.481 |
| Te | −0.252 | −0.059 | −0.267 | 1.086 | 1.309 | −0.332 | 1.551 | 1.104 |
| I | 0.694 | 0.440 | 0.149 | 2.320 | 1.714 | 0.871 | 2.162 | 1.491 |
| Cs | 1.018 | 0.050 | −0.053 | 2.851 | 1.494 | 1.056 | 2.000 | 1.713 |
| Ba | 0.592 | 0.054 | −0.044 | 1.091 | 0.056 | −0.187 | 0.608 | 0.577 |
| La | −1.094 | −0.152 | −1.234 | −0.476 | 0.562 | −0.805 | 0.500 | −0.447 |
| Ce | −0.847 | 0.061 | −1.049 | −1.080 | 0.191 | −0.916 | −0.121 | −0.819 |
| Pr | −0.732 | 0.133 | −0.957 | −1.137 | 0.103 | −0.904 | −0.222 | −0.883 |
| Nd | −0.662 | 0.181 | −0.883 | −1.105 | 0.088 | −0.883 | −0.214 | −0.875 |
| Pm | −0.613 | 0.205 | −0.821 | −1.032 | 0.091 | −0.855 | −0.170 | −0.840 |
| Sm | −0.581 | 0.213 | −0.775 | −0.947 | 0.103 | −0.830 | −0.111 | −0.797 |
| Eu | −0.571 | 0.206 | −0.760 | −0.913 | 0.130 | −0.809 | −0.098 | −0.755 |
| Gd | −0.572 | 0.199 | −0.766 | −0.845 | 0.154 | −0.805 | −0.049 | −0.736 |
| Tb | −0.582 | 0.198 | −0.767 | −0.827 | 0.186 | −0.797 | −0.005 | −0.713 |
| Dy | −0.586 | 0.190 | −0.763 | −0.778 | 0.209 | −0.788 | 0.041 | −0.686 |
| Ho | −0.586 | 0.185 | −0.749 | −0.737 | 0.229 | −0.776 | 0.078 | −0.653 |
| Er | −0.585 | 0.169 | −0.735 | −0.700 | 0.234 | −0.766 | 0.097 | −0.624 |
| Tm | −0.588 | 0.169 | −0.720 | −0.663 | 0.260 | −0.752 | 0.136 | −0.588 |
| Yb | −0.615 | 0.131 | −0.714 | −0.607 | 0.286 | −0.733 | 0.179 | −0.525 |
| Lu | −0.524 | 0.091 | −0.575 | −1.180 | −0.021 | −0.724 | −0.545 | −0.639 |
| Hf | −0.164 | 0.041 | −0.104 | −1.927 | −0.604 | −0.562 | −1.681 | −0.666 |
| Ta | 0.192 | 0.439 | 0.312 | −2.094 | −0.437 | −0.311 | −1.806 | −0.466 |
| W | 0.449 | 0.240 | 0.537 | −1.769 | −0.610 | −0.112 | −1.959 | −0.209 |
| Re | 0.615 | −0.093 | 0.502 | −1.332 | −0.593 | −0.182 | −2.033 | 0.116 |
| Os | 0.591 | −0.488 | 0.318 | −0.785 | −0.663 | −0.306 | −1.850 | 0.246 |
| Ir | 0.458 | −0.768 | 0.052 | −0.192 | −0.752 | 0.157 | −1.388 | 0.116 |
| Pt | 0.245 | −0.775 | −0.102 | 0.224 | −0.687 | −0.214 | −0.732 | −0.050 |
| Au | −0.064 | −0.507 | −0.207 | 0.359 | −0.243 | −0.132 | −0.015 | 0.020 |
| Hg | −0.511 | −0.405 | −0.399 | 0.509 | 0.444 | −0.178 | 0.651 | 0.490 |
| Tl | −0.817 | −0.389 | −0.695 | 0.938 | 0.921 | 0.175 | 1.455 | 0.658 |
| Pb | −0.899 | −0.390 | −0.786 | 0.924 | 0.890 | 0.072 | 1.552 | 0.573 |
| Bi | −0.464 | 0.019 | −0.373 | 0.923 | 0.928 | −0.395 | 1.606 | 0.649 |

**Table S2** Segregation Energies and Embrittlement Potencies from PAW method.

| Solute | Potential | $\Delta E_1^{seg}$ | $\Delta E_2^{seg}$ | $\Delta E_3^{seg}$ | $\Delta E_1^{emb}$ | $\Delta E_{2-I}^{emb}$ | $\Delta E_{3-I}^{emb}$ | $\Delta E_{2-II}^{emb}$ | $\Delta E_{3-II}^{emb}$ |
|---|---|---|---|---|---|---|---|---|---|
| Na | Na_sv | −0.662 | −0.002 | −0.644 | 0.780 | 0.948 | −0.025 | 1.487 | 0.312 |
| Mg | Mg_sv | −0.336 | 0.146 | −0.220 | −0.108 | 0.390 | −0.254 | 0.403 | 0.019 |
| Si | Si | 0.096 | −0.278 | −0.089 | 0.555 | −0.213 | 0.002 | 0.257 | 0.038 |
| P | P | −0.200 | −0.493 | −0.161 | 0.777 | 0.233 | −0.006 | 0.610 | 0.614 |
| S | S | −0.251 | −0.657 | −0.213 | 1.028 | 0.149 | −0.050 | 0.767 | 1.234 |
| Cl | Cl | 0.190 | 0.088 | 0.036 | 1.896 | 1.533 | −0.315 | 1.909 | 1.494 |
| K | K_sv | 0.430 | 1.258 | −0.038 | 1.964 | 2.240 | 0.654 | 2.794 | 0.887 |
| Ca | Ca_sv | −0.866 | −0.017 | −0.917 | 0.034 | 0.755 | −0.616 | 0.889 | −0.108 |
| Sc | Sc_sv | −0.292 | 0.117 | −0.290 | −1.122 | −0.165 | −0.571 | −0.702 | −0.529 |
| Ti | Ti_sv | 0.165 | 0.462 | 0.480 | −1.469 | −0.262 | −0.081 | −1.135 | −0.167 |
| V | V_sv | 0.445 | 0.305 | 0.543 | −1.356 | −0.501 | −0.115 | −1.459 | −0.166 |
| Cr | Cr_sv | 0.602 | 0.013 | 0.511 | −1.047 | −0.596 | −0.144 | −1.610 | 0.033 |
| Mn | Mn_sv | 0.608 | −0.388 | 0.409 | −0.790 | −0.626 | 0.138 | −1.770 | 0.302 |
| Fe | Fe_sv | 0.555 | −0.688 | 0.056 | −0.449 | −0.736 | 0.075 | −1.687 | 0.137 |
| Co | Co_pv | 0.458 | −0.850 | −0.118 | −0.123 | −0.842 | 0.051 | −1.397 | −0.022 |
| Ni | Ni_pv | 0.362 | −0.797 | −0.096 | 0.143 | −0.834 | 0.005 | −0.951 | −0.097 |
| Cu | Cu | 0.255 | −0.466 | −0.021 | 0.283 | −0.557 | −0.081 | −0.388 | −0.165 |
| Zn | Zn | −0.013 | −0.189 | −0.076 | 0.223 | −0.095 | −0.042 | 0.046 | −0.011 |
| Ga | Ga_h | −0.137 | −0.203 | −0.147 | 0.364 | 0.039 | −0.038 | 0.309 | 0.109 |
| Ge | Ge_h | −0.164 | −0.272 | −0.207 | 0.614 | 0.028 | −0.055 | 0.600 | 0.157 |
| As | As | −0.308 | −0.373 | −0.315 | 0.837 | 0.447 | −0.104 | 0.922 | 0.554 |
| Se | Se | −0.156 | −0.312 | −0.185 | 1.041 | 0.980 | −0.214 | 1.032 | 1.179 |
| Br | Br | −0.164 | −0.132 | −0.318 | 2.095 | 1.706 | −0.388 | 2.234 | 1.529 |
| Rb | Rb_sv | 0.673 | −0.059 | −0.179 | 2.638 | 1.370 | 0.930 | 1.885 | 1.149 |
| Sr | Sr_sv | −0.694 | 0.167 | −0.793 | 0.599 | 1.090 | −0.183 | 1.467 | 0.198 |
| Y | Y_sv | −0.725 | 0.038 | −0.806 | −0.998 | 0.163 | −0.790 | −0.235 | −0.626 |
| Zr | Zr_sv | −0.257 | 0.039 | −0.215 | −1.816 | −0.534 | −0.636 | −1.492 | −0.701 |
| Nb | Nb_sv | 0.146 | 0.415 | 0.316 | −1.896 | −0.399 | −0.307 | −1.602 | −0.400 |
| Mo | Mo_sv | 0.406 | 0.211 | 0.517 | −1.505 | −0.575 | −0.162 | −1.697 | −0.167 |
| Tc | Tc_sv | 0.577 | −0.090 | 0.476 | −1.077 | −0.612 | −0.223 | −1.735 | 0.073 |
| Ru | Ru_pv | 0.543 | −0.449 | 0.294 | −0.567 | −0.699 | 0.074 | −1.536 | 0.149 |
| Rh | Rh_pv | 0.418 | −0.624 | 0.080 | −0.074 | −0.762 | 0.020 | −1.065 | −0.001 |
| Pd | Pd | 0.221 | −0.510 | −0.040 | 0.208 | −0.583 | −0.171 | −0.430 | −0.153 |
| Ag | Ag_pv | −0.129 | −0.256 | −0.143 | 0.206 | −0.061 | −0.117 | 0.163 | 0.020 |
| Cd | Cd | −0.473 | −0.218 | −0.317 | 0.288 | 0.424 | −0.174 | 0.571 | 0.348 |
| In | In_d | −0.617 | −0.245 | −0.463 | 0.489 | 0.583 | −0.242 | 0.923 | 0.406 |
| Sn | Sn_d | −0.640 | −0.279 | −0.516 | 0.626 | 0.559 | −0.269 | 1.102 | 0.402 |
| Sb | Sb | −0.683 | −0.346 | −0.598 | 0.792 | 0.622 | −0.301 | 1.323 | 0.501 |
| Te | Te | −0.260 | −0.061 | −0.271 | 1.060 | 1.291 | −0.345 | 1.552 | 1.100 |
| I | I | 0.692 | 0.437 | 0.146 | 2.307 | 1.703 | 0.850 | 2.150 | 1.474 |
| Cs | Cs_sv | 1.024 | 0.058 | −0.058 | 3.226 | 1.623 | 1.271 | 2.223 | 1.802 |
| Ba | Ba_sv | 0.610 | 0.052 | −0.048 | 1.368 | 0.159 | −0.066 | 0.796 | 0.648 |
| La | La | −1.092 | −0.144 | −1.230 | −0.373 | 0.581 | −0.805 | 0.582 | −0.429 |
| Hf | Hf_pv | −0.169 | 0.042 | −0.109 | −1.910 | −0.601 | −0.561 | −1.664 | −0.664 |
| Ta | Ta_pv | 0.189 | 0.435 | 0.311 | −2.090 | −0.442 | −0.312 | −1.802 | −0.463 |
| W | W_sv | 0.449 | 0.235 | 0.533 | −1.770 | −0.623 | −0.118 | −1.961 | −0.213 |
| Re | Re_pv | 0.611 | −0.102 | 0.498 | −1.325 | −0.594 | −0.192 | −2.026 | 0.123 |
| Os | Os_pv | 0.586 | −0.500 | 0.313 | −0.780 | −0.664 | −0.314 | −1.847 | 0.252 |
| Ir | Ir | 0.454 | −0.773 | 0.049 | −0.195 | −0.754 | 0.163 | −1.389 | 0.116 |
| Pt | Pt_pv | 0.242 | −0.777 | −0.105 | 0.224 | −0.690 | −0.220 | −0.728 | −0.054 |
| Au | Au | −0.064 | −0.522 | −0.207 | 0.356 | −0.260 | −0.131 | −0.031 | 0.016 |
| Hg | Hg | −0.517 | −0.405 | −0.401 | 0.516 | 0.453 | −0.175 | 0.664 | 0.497 |
| Tl | Tl_d | −0.818 | −0.388 | −0.694 | 0.941 | 0.929 | 0.169 | 1.460 | 0.665 |
| Pb | Pb_d | −0.898 | −0.387 | −0.785 | 0.924 | 0.896 | 0.061 | 1.558 | 0.578 |
| Bi | Bi_d | −0.467 | 0.019 | −0.374 | 0.916 | 0.921 | −0.402 | 1.612 | 0.656 |